\documentclass[aps,prl,reprint,showpacs,superscriptaddress,citeautoscript]{revtex4-1}

\usepackage{lmodern}
\usepackage{amsmath}
\usepackage{amssymb}
\usepackage{textcomp}
\usepackage{gensymb}
\usepackage{graphicx}
\usepackage{wasysym}
\usepackage{units}
\usepackage{array}
\usepackage{multirow}

\usepackage[bookmarks=false]{hyperref}
\hypersetup{
		breaklinks=true,
    unicode=false,          
    pdftoolbar=true,        
    pdfmenubar=true,        
    pdffitwindow=false,     
    pdfstartview={FitH},    
    pdftitle={Taufour},    
    pdfauthor={Taufour},     
    pdfsubject={WingsEquations},   
    pdfcreator={Taufour},   
    pdfproducer={Taufour}, 
    pdfkeywords={keyword1} {key2} {key3}, 
    pdfnewwindow=true,      
    colorlinks=true,       	
    linkcolor=black,          
    citecolor=black,    	    
    filecolor=black,   		    
    urlcolor=blue            
}

\begin{document}

\title{Constraints on the merging of the transition lines at the tricritical point in a wing-structure phase diagram}
\author{Valentin \surname{Taufour}}
\email{vtaufour@ucdavis.edu}
\affiliation{The Ames Laboratory, US Department of Energy, Iowa State University, Ames, Iowa 50011, USA}
\author{Udhara S. \surname{Kaluarachchi}}
\affiliation{The Ames Laboratory, US Department of Energy, Iowa State University, Ames, Iowa 50011, USA}
\affiliation{Department of Physics and Astronomy, Iowa State University, Ames, Iowa 50011, USA}%
\author{Vladimir~G. \surname{Kogan}}
\affiliation{The Ames Laboratory, US Department of Energy, Iowa State University, Ames, Iowa 50011, USA}

\begin{abstract}
We consider the phase diagram of a ferromagnetic system driven to a quantum phase transition with a tuning parameter $p$. Before being suppressed, the transition becomes of the first order at a tricritical point, from which wings emerge under application of the magnetic field $H$ in the $T$-$p$-$H$ phase diagram. We show that the edge of the wings merge with tangent slopes at the tricritical point.
\end{abstract}

\maketitle

\begin{figure}[!htb]
\begin{center}
\includegraphics[width=8.0cm]{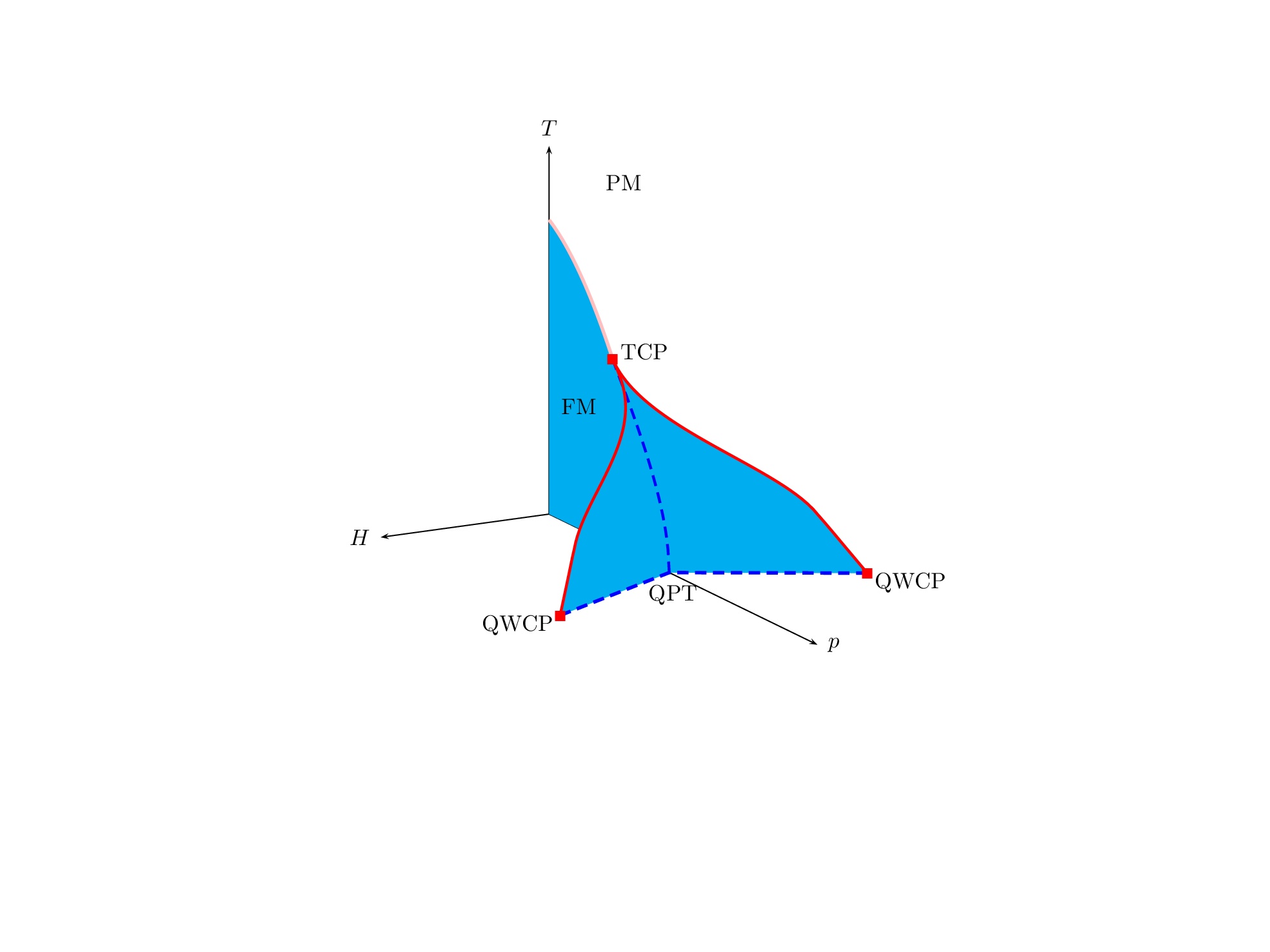}
\caption{(Color online) Schematic temperature-pressure-magnetic field ($T$-$p$-$H$) phase diagram. $H$ is applied along the easy axis of magnetization. The regions of the ferromagnetic (FM) and paramagnetic (PM) phases are shown. The diagram also indicates the tricritical point (TCP), the quantum phase transition lines (QPT), as well as two quantum wing critical points (QWCPs). Solid lines are second-order transitions. Dashed lines and blue planes are first-order transitions.\label{fig:diag3D}}
\end{center}
\end{figure}

The paramagnetic-ferromagnetic (PM-FM) phase transition is a textbook example of a second-order phase transition~\cite{LandauLifshitz5,KittelThermalPhysics,PLMphasetrans,Blundell2001Magnetism}. However, when suppressing the phase transition with a clean parameter such as pressure $p$, the transition becomes of the first order at the tricritical point (TCP) at $T_{\textrm{tcp}}$ and $p_{\textrm{tcp}}$~\cite{Brando2016RMP}. At larger pressures, $p_{\textrm{tcp}}<p<p_\textrm{qwcp}$, metamagnetic transitions can be observed when a magnetic field $H$ is applied along the direction of easy magnetization $M$. The typical $T$-$p$-$H$ phase diagram is illustrated in Fig.~\ref{fig:diag3D}. It is symmetric with respect to $H\rightarrow-H$, forming a so-called wing structure phase diagram. There are surfaces of first-order transitions (dark blue surfaces) which are limited by second-order transition lines (solid red lines~=~lines of critical points). These lines ($T_w$, $p_w$, $H_w$) start at the TCP and evolve towards $T=0$ at two quantum wing critical points (QWCPs)~\cite{Kirkpatrick2015PRB}. Such a three-dimensional phase diagram is often presented schematically~\cite{Pfleiderer2001Nature,Uhlarz2004PRL,Belitz2005PRL,Kimura2004PRL,Yamada2007PB,Rowley2010,Wu2011PRB,Kirkpatrick2012PRB,Kirkpatrick2015PRL,Brando2016RMP}. Recently, such a phase diagram was determined experimentally for a few compounds such as UGe$_2$~\cite{Taufour2010PRL,Kotegawa2011JPSJ} and ZrZn$_2$~\cite{Kabeya2012JPSJ}. As the current understanding of this phase diagram progresses, it was pointed out recently~\cite{Kirkpatrick2015PRL} that the Clapeyron relation~\cite{Clapeyron1834JEP} requires an infinite slope of the first-order transition at $T=0$. In addition, the wings are tilted in the direction of the disordered phase and are not perpendicular to the $p$ axis~\cite{Kirkpatrick2015PRL}. Here, we are interested in the behavior of the wing lines near the TCP. In particular, we show that the slope is infinite along the field axis, but finite and tangent to the other lines along the pressure axis. These results are based on generic thermodynamic arguments, but were not reflected in the phase diagrams drawn so far. The resulting schematic phase diagram is shown in Fig.~\ref{fig:diag3D}.

In condensed matter physics, phase diagrams are important tools as they provide maps of the existing phases and transitions. The experimental determination of phase diagrams can be difficult because of various unavoidable complications. Often, portions of the phase diagram are interpolated between the available experimental data. As is well known for constitutional binary diagrams used in material science, various thermodynamic phase rules apply to these diagrams. These rules allow one to improve the accuracy of phase diagrams and to detect or prevent possible errors~\cite{Okamoto1991JPE,Okamoto1993JPE,Okamoto1994JPE}. In the three-dimensional wing-structure phase diagram considered here, we find that the slopes of the phase transition lines must be tangent at the tricritical point in all three dimensions (Fig.~\ref{fig:diag3D}).

In order to reproduce a first-order PM-FM phase transition, the mean-field Landau expansion of the free energy must be extended to the sixth order~\cite{KittelThermalPhysics,LandauLifshitz5}:
\begin{eqnarray}
F=F_0+AM^2+BM^4+CM^6-HM\label{eqn:landau}
\end{eqnarray}
In Eq.~(\ref{eqn:landau}), $M$ is the order parameter and the coefficients $A$, $B$, and $C$ are functions of the temperature $T$ and pressure $p$.
The stable solutions correspond to $dF/dM=0$ and $d^2F/dM^2>0$. When the PM-FM transition is of the first order, there will be two degenerate solutions ($M_1$ and $M_2$) and an unstable solution for $M_1<M<M_2$ ($dF/dM=0$ and $d^2F/dM^2<0$). This means that $d^2F/dM^2$ has a minimum between $M_1$ and $M_2$ where $d^3F/dM^3=0$. The discontinuity at the first-order transition is $\Delta M=M_2-M_1$. As a function of $T$, $p$, or $H$, the first-order transition terminates when $M_1=M_2$. This corresponds to the edge of the wing (the line of critical points joining TCP and QWCP). At this line, $dF/dM=d^2F/dM^2=d^3F/dM^3=0$:
\begin{eqnarray}
\dfrac{dF}{dM}=&2AM_w+4BM_w^3+6CM_w^5-H_w&=0\label{eqn:dFdM0}\\
\dfrac{d^2F}{dM^2}=&2A+12BM_w^2+30CM_w^4&=0\label{eqn:d2FdM20}\\
\dfrac{d^3F}{dM^3}=&24M_w(B+5CM_w^2)&=0\label{eqn:d3FdM30}
\end{eqnarray}
The solution $M_w=0$ of Eq.~(\ref{eqn:d3FdM30}) is compatible with Eqs.~(\ref{eqn:dFdM0}) and (\ref{eqn:d2FdM20}) if $H_w=0$ and $A=0$. It corresponds to the tricritical point.

\begin{figure*}[!htb]
\begin{center}
\includegraphics[width=15cm]{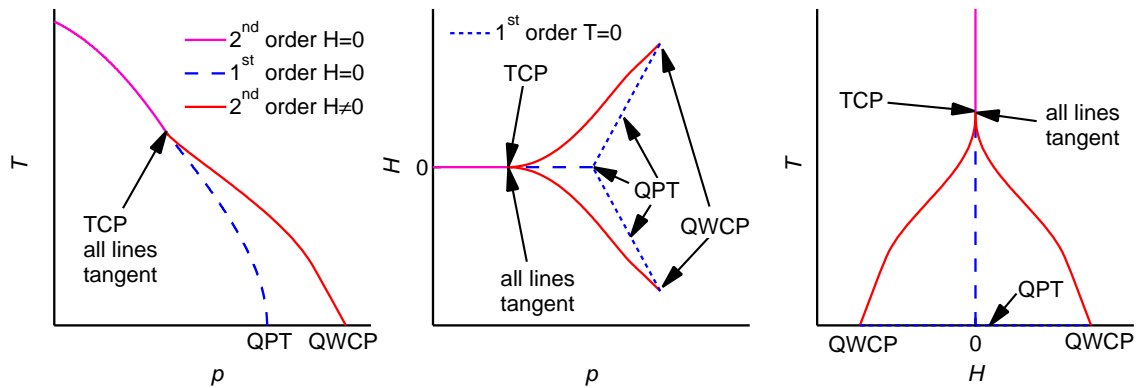}
\caption{(Color online) Schematic projections of the temperature-pressure-magnetic field ($T$-$p$-$H$) phase diagram in the $T$-$p$ ($H=0$), $H$-$p$ ($T=0$), and $T$-$H$ ($p=0$) planes.\label{fig:diagabc}}
\end{center}
\end{figure*}

We consider now the other solution of Eq.~(\ref{eqn:d3FdM30}):
\begin{eqnarray}
M_w^2=-\dfrac{B}{5C}\label{eqn:W1}
\end{eqnarray}
Then, Eq.~(\ref{eqn:d2FdM20}) yields:
\begin{eqnarray}
5AC-3B^2=0\label{eqn:W2}
\end{eqnarray}
Equation~(\ref{eqn:dFdM0}) then gives:
\begin{eqnarray}
2M_w(25AC-7B^2)&=&25CH_w\label{eqn:Hinterm}
\end{eqnarray}
Using Eq.~(\ref{eqn:W2}) with Eq.~(\ref{eqn:Hinterm}), we obtain:
\begin{eqnarray}
16M_wB^2&=&25CH_w\label{eqn:W3}
\end{eqnarray}
Equations~(\ref{eqn:W1}),~(\ref{eqn:W2}), and (\ref{eqn:W3}) determine the shape of the wing line in the $T$-$p$-$H$ phase diagram.

The slope along the wing line is obtained by differentiating~(\ref{eqn:W2}) and (\ref{eqn:W3}):
\begin{eqnarray}
5(CdA+AdC)-6BdB=0\label{eqn:S1}\\
16(B^2dM+2MBdB)=25(CdH+HdC)\label{eqn:S2}
\end{eqnarray}
As shown above, at the tricritical point, we have $A=B=H=M=0$, so that, at the TCP, Eqs.~(\ref{eqn:S1}) and (\ref{eqn:S2}) become:
\begin{eqnarray}
dA=0\\
dH=0
\end{eqnarray}
The relation $dA=0$ is also the equation for the slope of the second order phase transition line ($H=0$, $p<p_{\textrm{TCP}}$) and the slope of the first order transition line at TCP ($H=0$, $p=p_{\textrm{TCP}}$)~\cite{LandauLifshitz5}. Since $A$ depends only on $T$ and $p$, the equation~$dA=0$ implies that $dT/dp$ is the same for both lines at the TCP~\cite{LandauLifshitz5}. Having the same equation for the slope of the wing line means that the four lines (two wings and one second-order line and one-first order line) have the same tangent at the TCP. Hence, the extra dimension of $H$ in the $T$-$p$-$H$ phase diagram does not change the statement that the first- and second-order lines meet at the TCP without a kink. Hence, we have at the TCP:
\begin{eqnarray}
\dfrac{dT_w}{dp_w}=\dfrac{dT_{1\textrm{st}}}{dp_{1\textrm{st}}}=\dfrac{dT_{2\textrm{nd}}}{dp_{2\textrm{nd}}}
\end{eqnarray}
We note that $dH=0$ imposes that $\dfrac{dT_w}{dH_w}$ and $\dfrac{dp_w}{dH_w}$ are infinite. In other words, the wing lines approach the TCP being vertical.

The schematic diagram in Fig.~\ref{fig:diag3D} illustrates this result: all the lines are tangent to each other at the tricritical point. For a clearer view, Fig.~\ref{fig:diagabc} shows projections of such a three-dimensional diagram onto three planes. Our argument leading to the fact that the lines are tangent at TCP is completely general within the framework of mean-field theories. The exact shape of the phase diagram can vary substantially with the variation of the parameters $A$, $B$, and $C$ which are functions of $T$ and $p$. However, our demonstration does not use any restrictions on these parameters.

We note that the correct behavior of the lines is obtained in microscopic models for the origin of the first-order transition and wing-structure phase diagram~\cite{Belitz2005PRL,Wysokinski2015PRB,Abram2016JMMM}. It is not a surprise since renormalized Landau theories can be simplified as a Landau expansion to the sixth order~\cite{[{For example a non-analytic term such as $M^2\textrm{ln}(M^2+t^2)$ will reduce to $M^4\textrm{ln}t^2+M^6/t^2+\circ(M^7)$ when $t>0$}]EmptyBib}. However, experimentally, the wing lines are determined in a discrete manner: at a given pressure $p$, magnetic field (or temperature) sweeps are performed at discrete temperature (or magnetic field) values. The wing lines are interpolated between the experimental points so that the tangent merging of the lines at the TCP can be overlooked~\cite{Taufour2010PRL,Kotegawa2011JPSJ,Kabeya2012JPSJ}.

Our results have implications for a precise determination of the position of the tricritical point. For UGe$_2$, a first-order transition with a field induced recovery of a second-order-like anomaly was observed at $1.46$~GPa, whereas no such behavior was observed at $1.37$~GPa~\cite{Taufour2010PRL}. Hence, the position of the tricritical point was given as $p_{\textrm{tcp}}\approx1.42$~GPa. This value was also consistent with a linear extrapolation of the wing line to $H=0$, whereas, according to the present study, a linear slope in this region is not physical. The TCP is thus at lower pressures than $1.42$~GPa. In fact, the magnetic field step used at $1.37$~GPa being $0.2$~T, one can only say that, if the TCP is under $1.37$~GPa, the wing-line is below $0.2$~T. Interestingly, a recent microscopic modeling of the wing-lines based on the Anderson lattice model was proposed~\cite{Wysokinski2015PRB,Abram2016JMMM}, in which the position of the tricritical point was imposed at $1.42$~GPa. Perhaps a better agreement with the experiments can be obtained if the TCP is at lower pressures with the condition that the wing line at $1.37$~GPa ($24.5$~K) is below $0.2$~T.

In several reports, the term quantum critical end point was misused to label the point at which the wing lines reach $T=0$~K~\cite{Pfleiderer2001Nature,Grigera2001Science,Millis2002PRL,Grigera2003PRB,Taufour2010PRL,Kotegawa2011JPSJ,Wu2011PRB}. This was pointed out recently in Refs.~\cite{Kirkpatrick2015PRB,Brando2016RMP} and the name quantum wing critical point was proposed~\cite{Kirkpatrick2015PRB}. One important difference with a quantum critical point is that there is no spontaneous symmetry breaking at the QWCP~\cite{Grigera2001Science}. In some situations, a Lifshitz transition can occur simultaneously with the quantum phase transition. Such a change of Fermi surface topology at $T=0$~K cannot be achieved continuously via a crossover around the QWCP. This means that a quantum critical line continues to exist at $T=0$~K at larger magnetic field and pressure. The term marginal quantum critical point has been used in this case~\cite{Yamaji2006JPSJ,Yamaji2007JPSJ,Kabeya2012JPSJ}.

In conclusion, we have shown that all transition lines merging at a tricritical point are tangent. This was known for the first and second order lines at $H=0$~\cite{LandauLifshitz5} and our results extend it to the three-dimensional $T$-$p$-$H$ space. In particular, the wing-lines, that emerge in a three-dimensional phase diagram, have a tangent slope at the tricritical point. This result implies that very low field measurements need to be performed in order to precisely determine the position of the tricritical point. We hope that our work will help improve the accuracy of determination of phase diagrams.

We would like to thank D.~K.~Finnemore, S.~L.~Bud'ko, P.~C.~Canfield, and A.~E.~B\"ohmer for useful discussions and P.~C.~Canfield for critical review of the manuscript.
This work was supported by the Materials Sciences Division of the Office of Basic Energy Sciences of the U.S. Department of Energy. This work was performed at the Ames Laboratory, U.S. DOE, under Contract No. DE-AC02-07CH11358.

\end{document}